\documentclass[reprint, amsmath,amssymb,aps, preprintnumbers, nofootinbib]{revtex4-1}
\usepackage{graphicx}				
\usepackage{dcolumn}
\usepackage{bm}
\usepackage{slashed}
\usepackage{tikz}
\usepackage{rotating}
\usepackage{hyperref}
\usepackage{array}
\usepackage{color}
\usepackage{multirow}

\begin{document}
\preprint{DESY 18-034}

\title{Pocket Formulae for Non-Abelian Discrete Anomaly Freedom}

\author{Jim Talbert}
\email{james.talbert@desy.de}
\affiliation{Theory Group, Deutsches Elektronen-Synchrotron (DESY), D-22607 Hamburg, Germany}

\begin{abstract}
We show that the discrete anomaly constraints governing popular non-Abelian symmetries of use in (e.g.) flavoured, supersymmetric, and dark matter model building typically subdivide into two classes differentiated by the simple restrictions they impose on the \emph{number} of fields transforming under certain irreducible representations of the relevant groups.  These constraints lead us both to generic conclusions for common Beyond-the-Standard-Model constructions (including rather powerful statements for Grand Unified theories) as well as to simplified formulae that can be rapidly applied to determine whether a given field and symmetry content suffers from gauge and gravitational anomalies.
\end{abstract}

\maketitle

\section{Introduction}
\label{sec:intro}
Discrete symmetries are ubiquitous in Beyond-the-Standard-Model (BSM) constructions. Not only are they utilized ad-hoc to prevent unwanted couplings, as is often required to (e.g.) stabilize a dark matter candidate, they also have more purposeful implementations; non-Abelian discrete symmetries can explain observed patterns of fermionic mass and mixing (see e.g. \cite{Altarelli:2010gt,King:2013eh}), control models of inflation \cite{Ross:2009hg}, and may even be naturally realized as interchange symmetries of fixed-points in orbifold compactifications \cite{Kobayashi:2006wq,Adulpravitchai:2009id}. 

Regardless of motivation, imposing a discrete symmetry on a fixed Lagrangian practically amounts to manipulating a global symmetry --- no additional gauge bosons are present.  However, it has long been argued that global discrete symmetries \emph{must} be gauged in the ultra-violet (UV) in order to respect quantum gravity (wormhole) effects  \cite{Krauss:1988zc,Gilbert:1989nq,Choi:2017xmg},
and  therefore models employing discrete symmetries should be anomaly free.  Constraints for Abelian discrete symmetries were first obtained in \cite{Ibanez:1991hv,Ibanez:1991wt,Banks:1991xj} by assuming that cyclic $Z_{N}$ groups (with N the order of the group) originate from the breakdown of a gauged $U(1)$.  Analogous considerations were made for non-Abelian discrete symmetries in \cite{Frampton:1994rk,Luhn:2008sa}.  These studies have since been generalized \cite{Araki:2007zza,Araki:2008ek,Ishimori:2010au,Chen:2015aba} \footnote{We largely follow the notation of \cite{Chen:2015aba} in the equations that follow.} with a path-integral approach \cite{Fujikawa:1979ay,Fujikawa:1980eg}, with the conclusion that a fully massless spectrum in the IR is only subject to mixed non-Abelian gauge ($G$) and gravitational ($\mathcal{G}$) anomaly constraints of the form:
\begin{equation}
D-G-G, \,\,\,\,\,\,\,\,\,\, D-\mathcal{G}-\mathcal{G}
\end{equation}
where $D$ can be either an Abelian or non-Abelian discrete symmetry.  Triangles like $\left[D \right]^{2} U(1)$ and $\left[ U(1)\right]^{2} D$ do not provide concrete information in the IR because the corresponding discrete charge $\alpha$ of any group element transformation is always defined modulo $N$.  One can always rescale the hypercharges of the $U(1)$ symmetry groups to satisfy this modulo constraint.  Also, cubic discrete anomalies ($\left[D\right]^{3}$) can be avoided by arguing charge fractionalization in the massive particle spectrum \cite{Ibanez:1991hv,Ibanez:1991wt,Banks:1991xj,Csaki:1997aw},\footnote{Taking the charge fractionalization approach, anomalies following from cubic constraints can give valuable information about the ultimate order required of the $D$ groups for the model to be completely consistent.  I thank G.G. Ross for this comment.} and indeed do not even appear in the path integral approach \cite{Araki:2008ek,Lee:2011dya}.

In this note we extend \cite{Araki:2007zza,Araki:2008ek,Ishimori:2010au,Chen:2015aba} by showing that, after reorganizing discrete anomaly constraints into compact multiplicative forms, popular non-Abelian discrete symmetries are generally subject to one of two classes of constraints distinguished by the restriction they impose on the \emph{number} of fields transforming in certain irreducible representations (irreps) of $G$ and $D$.  This leads us to a host of generic conclusions that are relevant in many BSM contexts, especially Grand Unified theories, and which at the very least yield simplified `field equations' that can be rapidly applied to concrete models.

The paper develops as follows:  in Section \ref{sec:formalism} we review the path-integral formalism developed in \cite{Araki:2007zza,Araki:2008ek,Ishimori:2010au,Chen:2015aba}, extend it to obtain the new multiplicative basis, and then specify the resulting anomaly constraints to explicit discrete symmetry groups. Then in Section \ref{sec:symmetries} we place said constraints into two classes, each characterized by a `field equation,' before deriving generic conclusions and simplified formulae relevant to common model building scenarios, examples of which we explore in Section \ref{sec:APPS}. We conclude in Section \ref{sec:conclusions}.
\section{Discrete Anomaly Constraints}
\label{sec:formalism}
Consider a set of Dirac fermions $\Psi$ living in the irreps $\bold{r}$ and $\bold{d}$ of a non-Abelian gauge $G$ and non-Abelian discrete group $D$, respectively.  Then the unitary representation of a discrete transformation associated to the element $g \in D$ is given by
\begin{equation}
\label{eq:unitrep}
U_{\bold{d}}(g) =  e^{i \alpha_{\bold{d}}(g)} = e^{i \,2 \pi \,\tau_{\bold{d}}(g)/N_{g}}
\end{equation}
in terms of a charge $\alpha_{\bold{d}}$ defined by $N_{g}$, the order of the element $g$, and a charge matrix $\tau_{\bold{d}}(g)$ which has integer eigenvalues.  In general, chiral transformations of the fermions $\Psi$ under $U_{\bold{d}}(g)$ source a Jacobian in the path integral measure of a quantum field theory \cite{Fujikawa:1979ay,Fujikawa:1980eg}: 
\begin{equation}
\mathcal{D}\Psi \mathcal{D} \bar{\Psi} \underset{U} {\longrightarrow} J^{-2}(\alpha(g)) \mathcal{D}\Psi \mathcal{D} \bar{\Psi} 
\end{equation} 
where, if the Jacobian is found to be non-trivial ($J \neq 1$), the symmetry $D$ is \emph{anomalous}.  

There are both gauge $G$ and gravitational $\mathcal{G}$ contributions to the anomaly \cite{AlvarezGaume:1983ig,AlvarezGaume:1984dr,Fujikawa:1986hk}, but we focus on the former for the moment.  Consider the Jacobian for transformations on left-handed fields, $\psi_{L} \rightarrow \psi^{\prime}_{L} = e^{i \alpha_{\bold{d}}(g)}\psi_{L}$:
\begin{equation}
\label{Jacobian}
J^{-2}_{G} = exp \left( i \int d^{4}x \, \frac{1}{16 \pi^{2}} \text{tr} \left[ \alpha_{\bold{d}}(g)F^{\mu \nu} \tilde{F}_{\mu \nu} \right] \right)
\end{equation} 
where the trace runs over all internal indices and the field strength tensor embeds the generators $t$ of the associated gauge group $G$, $F_{\mu \nu} = F_{\mu \nu}^{a} t_{a}(\bold{r})$.  Its dual is given by $\tilde{F} ^{\mu \nu} = \frac{1}{2} \epsilon^{\mu \nu \rho\sigma} F_{\rho \sigma}$.  Recalling the index theorems of \cite{AlvarezGaume:1983ig,AlvarezGaume:1984dr} and defining the Dynkin index of the gauge representation $\bold{r}$ by\begin{equation}
\label{Dynkin}
l (\bold{r}) \delta_{ab} = tr \left[ t_{a} (\bold{r}) t_{b} (\bold{r})\right]
\end{equation}
we define the function $p$ as 
\begin{equation}
\label{index}
p \equiv \int d^{4}x \frac{1}{64 \pi^{2}} \epsilon^{\mu \nu \rho \sigma} F^{a}_{\mu \nu} F^{a}_{\rho \sigma} \,\,\, \in \,\,\, \mathbb{Z}
\end{equation}
(with $\mathbb{Z}$ denoting `integers') and then observe that \eqref{Jacobian} reduces to
\begin{equation}
J^{-2}_{G} = exp \left( i \frac{2 \pi}{N_{g}} \cdot \text{tr}\left[\tau_{\bold{d}}(g) \right] \cdot 2 l(\bold{r}) \cdot p \right)
\end{equation}
such that the transformation $U_{\bold{d}}(g)$ is free of $D-G-G$ anomalies if and only if \cite{Araki:2007zza,Araki:2008ek,Ishimori:2010au,Chen:2015aba}:
\begin{equation}
\label{zanom}
\underset{f}{\sum}\, \text{tr}\left[\tau_{\bold{d}^{(f)}}(g)\right] \cdot l(\bold{r}^{(f)}) \overset{!}{=} 0 \,\, \text{mod}\,\, \frac{N_{g}}{2}
\end{equation}
where the notation in \eqref{zanom} implies that the summation is only over chiral fermions $f$ living in representations that are non-trivial with respect to both $G$ and $D$.  Here it is clear that $\text{tr}\left[\tau_{\bold{d}^{(f)}}(g)\right]$ mimics a $Z_{N_{g}}$ charge that can be written in terms of a (multi-valued) logarithm:
\begin{equation}
\label{NAcharge}
\text{tr}\left[\tau_{\bold{d}^{(f)}}(g)\right] = N_{g} \frac{\text{ln}\, \text{det}\left[ U_{\bold{d}^{(f)}}(g)\right]}{2 \pi i}
\end{equation}
Were we to repeat the above analysis for $D-\mathcal{G}-\mathcal{G}$ triangles, we would find the following constraint  \cite{Araki:2008ek,Ishimori:2010au}: 
\begin{equation}
\label{eq:Ganom}
\underset{f}{\sum}\, \text{tr}\left[\tau_{\bold{d}^{(f)}}(g)\right] \overset{!}{=} 0 \,\, \text{mod}\,\, \frac{N_{g}}{2}
\end{equation} 
where now it is understood that the summation is over all chiral fermions non-trivial in $D$, irrespective of $G$.\footnote{Note that in both \eqref{zanom} and \eqref{eq:Ganom} additional gauge symmetry factors are left implicit.} 

We conclude that gauge and gravitationally anomalous transformations correspond to those with $\text{det}\left[ U_{\bold{d}^{(f)}}(g)\right] \neq 1$, a condition that must be checked for all $g \in D$.

\subsection{The Multiplicative Approach}
\label{sec:Jacobian}
As observed in \cite{Chen:2015aba}, one can rewrite the Jacobian using \eqref{NAcharge} 
to find multiplicative anomaly constraints (here written only for $D-G-G$):
\begin{equation}
\label{constraint2}
\underset{f}{\prod} \text{det} \left[ U_{\bold{d}^{(f)}}(g)\right]^{2\, l(\bold{r}^{(f)})} \overset{!}{=} 1
\end{equation}
Like its additive equivalent \eqref{zanom}, \eqref{constraint2} must be checked for every element $g \in D$.  At least two simplifying approaches exist in the literature to do so:
\begin{enumerate}
\item  Imagine that $D$ is generated by two elements $\lbrace h_{1}, h_{2} \rbrace \in D$, and that we have found that the Abelian transformations represented by $\lbrace h_{1}, h_{2} \rbrace$ are anomaly free.  As any other element $g \in D$ can be seen as a product of $h_{1}$ and/or $h_{2}$, this implies that $g$ is itself anomaly free.  Therefore, calculating \eqref{zanom} for each generator $h_{i}$ of $D$ is sufficient to determine anomaly freedom \cite{Araki:2007zza,Araki:2008ek}.  
\item Finite groups are also subdivided into conjugacy classes $C_{i}$, such that two elements $g_{1,2}$ belong to the same $C_{i}$ if and only if they are related by conjugation:  $g g_{1} g^{-1} = g_{2}$ for an element $g \in D$.  Since $\text{det}(gg_{i}g^{-1}) =\text{det}(g_{i})$, the determinant is constant over a conjugacy class and it is therefore sufficient to calculate \eqref{zanom} for each $C_{i}$ \cite{Ishimori:2010au}. 
\end{enumerate}

While these arguments were made in the context of \eqref{zanom}, they are also true for \eqref{constraint2}.  That is, the constraints
\begin{align}
\label{eq:GENconstraint2}
\underset{f}{\prod} \text{det} \left[ U_{\bold{d}^{(f)}}(h_{i})\right]^{(2\, l(\bold{r}^{(f)}),\,2)} &\overset{!}{=} 1 \\
\label{eq:CJconstraint2}
\underset{f}{\prod} \text{det} \left[ U_{\bold{d}^{(f)}}(C_{i})\right]^{(2\, l(\bold{r}^{(f)}),\,2)} &\overset{!}{=} 1
\end{align}
represent equivalent approaches to determining $D-(G,\mathcal{G})-(G,\mathcal{G})$ anomaly freedom.  In either case, one requires the determinants over either $h$ or $C$ for each irrep of $D$.

We now observe that the left-hand-sides (LHS) of \eqref{eq:GENconstraint2}-\eqref{eq:CJconstraint2} will always be composed of a finite number of basis elements, generically denoted $x^{a_{i}}_{i}$, with both $x$ and $a$ implicitly depending on the irrep $\bold{d}^{(f)}$, and $a$ also depending on the gauge representations $\bold{r}^{(f)}$.   That is,
\begin{equation}
\label{eq:basisone}
x_{1}^{a_{1}} \cdot x_{2}^{a_{2}} \,\text{...}\, x_{M-1}^{a_{M-1}} \cdot x_{M}^{a_{M}} \overset{!}{=} 1
\end{equation}
where $M$ represents the number of irreps of $D$.\footnote{The determinants $\text{det} \left[ U_{\bold{d}^{(f)}}(h_{i})\right]$ and $\text{det} \left[ U_{\bold{d}^{(f)}}(C_{i})\right]$ are one-dimensional and hence the basis elements $x_{i}$ can always be rescaled to a common multiple $x$, $x_{i}^{s_{i}} = x$, such that
\begin{equation}
\label{eq:basistwo}
x_{1}^{a_{1}} \cdot x_{2}^{a_{2}} \,\text{...}\, x_{M-1}^{a_{M-1}} \cdot x_{M}^{a_{M}} \equiv  x^{A_{(\bold{d},\bold{r})}} 
\end{equation}
from which one can rederive the constraints catalogued in \cite{Ishimori:2010au} by identifying the order $\mathcal{N}$ of the basis element $x$ and scale factors $s_{i}$ required to obtain $A_{(\bold{d},\bold{r})}$.} Restoring the dependence on the irreps,  we derive alternative anomaly constraints from \eqref{eq:basisone}:
\begin{align}
\label{eq:basisthree}
B^{(G,\mathcal{G})}_{(\bold{d},\bold{r})} \equiv \sum_{\bold{d}^{(f)}} \, a^{(G,\mathcal{G})}_{(\bold{d}^{(f)}, \bold{r}^{(f)})} \,\ln x_{\bold{d}^{(f)}} &\overset{!}{=} 0 \,\, \text{mod} \,\,2 \pi i 
\end{align} 
The functions $a^{(G,\mathcal{G})}_{(\bold{d}^{(f)}, \bold{r}^{(f)})}$ depend on the number of fields transforming in irreps of  $G$ and $D$ as well as any additional gauge symmetry factors, and will obviously differ between $D-G/\mathcal{G}-G/\mathcal{G}$ calculations.  The basis elements $x_{\bold{d}}$ are fixed numbers that can be extracted from any finite group. Indeed, in \eqref{eq:basisthree} the dependence on the (normalized) charge of any fermion under $D$ is fully factorized, a fact we will exploit in the next section.  

Although computable with mathematics software equipped with finite group libraries, we now perform explicit extractions of the basis logarithms $\ln x_{\bold{d}}$ from the $D_{N}$ and $\Delta(3N^{2})$ series while leaving other groups to Table \ref{tab:two}.\footnote{In what follows we use the catalogue in \cite{Ishimori:2010au} and maintain their notation on irreps.}

\subsubsection{$D_{N}$}
\label{sec:DN}

The dihedral groups $D_{N}$ describe the symmetries of $N$-sided regular polygons
and are composed of $Z_{N}$ cyclic rotations and $Z_{2}$ reflections; they are isomorphic to $Z_{N} \rtimes Z_{2}$.  $D_{N}$ is order $2N$ and is generated by two elements.  For odd $N$ the group has $(N+3)/2$ conjugacy classes and irreps, whereas for even $N$ there are $3+N/2$.  Dihedral groups have applications in flavoured \cite{Grimus:2003kq,Blum:2007jz,Hagedorn:2012pg,Varzielas:2016zuo}, inflationary \cite{Ross:2009hg}, and dark matter \cite{Adulpravitchai:2011ei} model-building.  From Table \ref{tab:3N2} one observes that for even $N$ we obtain $\lbrace (\bold{d}, \ln x_{\bold{d}}/(\pi i)) \rbrace_{a} = \lbrace (\bold{1_{-+}}, 1), (\bold{1_{--}}, 1), (\bold{2_{k}}, 1) \rbrace$ under transformations of the generator $a$ and $\lbrace (\bold{d}, \ln x_{\bold{d}}/(\pi i)) \rbrace_{ab} = \lbrace (\bold{1_{+-}}, 1), (\bold{1_{--}}, 1), (\bold{2_{k}}, 1) \rbrace$ under $ab$.  In both cases $k \in \lbrace 1, (N-2)/2 \rbrace$.  
For odd $N$ we are only concerned with transformations under the generator $b$, $\lbrace (\bold{d}, \ln x_{\bold{d}}/(\pi i)) \rbrace_{b} = \lbrace (\bold{1_{-}}, 1), (\bold{2_{k}}, 1) \rbrace$ with $k \in \lbrace 1, (N-1)/2 \rbrace$.

\begin{table}[t]
\renewcommand{\arraystretch}{1.25}
\noindent
\centering
\begin{tabular}{|c|c|c|c|}
\hline 
\multicolumn{3}{|c|}{$\Delta(3N^{2})$, $N/3 \in \mathbb{Z}$} \\
\hline
& $\bold{1_{k,l}}$ & $\bold{3_{[k][l]}}$ \\
\hline
det$\left( b \right)$  & $\omega_{3}^{k}$ & 1  \\
\hline
det$\left( a \right)$  & $\omega_{3}^{l}$ & 1 \\
\hline
det$\left( a^{\prime} \right)$  & $\omega_{3}^{l}$ & 1 \\
\hline
 \end{tabular}
 \,\,\,\,
 \begin{tabular}{|c|c|c|}
 \hline 
\multicolumn{3}{|c|}{$\Delta(3N^{2})$, $N/3 \notin \mathbb{Z}$} \\
\hline
& $\bold{1_{k}}$ & $\bold{3_{[k][l]}}$ \\
\hline
det$\left( b \right)$  & $\omega_{3}^{k}$ & 1  \\
\hline
det$\left( a \right)$  & 1 & 1 \\
\hline
det$\left( a^{\prime} \right)$  & 1 & 1 \\
\hline
\end{tabular} \\ \,\,\,\,\,\,
\begin{tabular}{|c|c|c|c|}
\hline 
\multicolumn{4}{|c|}{$D_{N}$, $N$ odd} \\
\hline
 & $\bold{1_{+}}$ & $\bold{1_{-}}$ & $\bold{2_{k}}$  \\
\hline 
det$\left( b \right)$  & 1 & -1 & -1 \\
\hline
 det$\left( a \right)$   & 1 & 1 & 1 \\
 \hline
 \end{tabular}
 \,\,\,\,
 \begin{tabular}{|c|c|c|c|c|c|}
 \hline 
\multicolumn{6}{|c|}{$D_{N}$, $N$ even} \\
\hline
& $\bold{1_{++}}$ & $\bold{1_{+-}}$ &$\bold{1_{-+}}$ & $\bold{1_{--}}$ & $\bold{2_{k}}$ \\
\hline
det$\left( a \right)$  & 1 & 1 & -1 & -1 & -1  \\
\hline
det$\left( a b \right)$  & 1 & -1 & 1 & -1 & -1 \\
\hline
\end{tabular}

\caption{TOP:  The determinants over the generators and irreps of $\Delta(3N^{2})$.  BOTTOM:  The same for $D_{N}$.}

\label{tab:3N2}
\end{table}
\subsubsection{$\Delta(3N^{2})$}
\label{sec:3N2}
The series $\Delta(3N^{2})$ is known (along with $\Delta(6N^{2})$) to have realistic applications in flavoured model building \cite{deMedeirosVarzielas:2006fc,Ma:2006ip,Varzielas:2015aua,Luhn:2007uq,deMedeirosVarzielas:2017sdv,Lam:2012ga,Holthausen:2012wt,King:2013vna,Holthausen:2013vba,Ishimori:2014jwa,Lavoura:2014kwa,Talbert:2014bda,Ishimori:2014nxa,Yao:2015dwa}.  The group is isomorphic to $\left(Z_{N} \times Z^{\prime}_{N} \right) \rtimes Z_{3}$ and can be generated by the three elements $\lbrace a, a^{\prime}, b \rbrace$ associated to $Z_{N,N^{\prime},3}$ respectively.  The group is order $3N^{2}$, and when $N/3 \notin \mathbb{Z}$ there are three singlet and $(N^{2} - 1)/3$ three-dimensional representations, whereas when $N/3 \in \mathbb{Z}$ there are nine singlet and $(N^{2} - 3)/3$ three-dimensional representations.  The determinants over the generators in these irreps are given in Table \ref{tab:3N2}.  Note that the tetrahedral group $A_{4}$, which is useful in dark matter \cite{Hirsch:2010ru} and flavoured \cite{Babu:2002dz,Ma:2002yp,Ma:2001dn,Altarelli:2005yx} model building, is isomorphic to $\Delta(12)$.

For $N/3 \in \mathbb{Z}$ there are potentially anomalous transformations under all three generators $a$, $a^{\prime}$, and $b$ when fermions sit in the singlet representations.  However, one notices that there are only two independent parameter sets: $\lbrace (\bold{d}, \ln x_{\bold{d}}/(\pi i)) \rbrace_{a} = \lbrace  (\bold{1_{k,l}}, 2l/3) \rbrace$ and $\lbrace (\bold{d}, \ln x_{\bold{d}}/(\pi i))  \rbrace_{b} = \lbrace(\bold{1_{k,l}}, 2k/3) \rbrace$.  
Whenever $N/3 \notin \mathbb{Z}$, the only irrep that contributes is $\bold{1_{k}}$ for $k\neq 0$, yielding $\lbrace (\bold{d}, \ln x_{\bold{d}}/(\pi i)) \rbrace_{b} = \lbrace  (\bold{1_{k}}, 2k/3) \rbrace$.

\section{Simplified Anomaly Constraints}
\label{sec:symmetries}

\begin{table}
\renewcommand{\arraystretch}{1.5}
\noindent
\centering
\begin{tabular}{|c|c|}
\hline 
\multicolumn{2}{|c|}{Basis Logarithms} \\
\hline
 Group &  $ \lbrace (\bold{d},\ln x_{\bold{d}}/(\pi i)) \rbrace$  \\
\hline 
\hline
 $D_{N \in odd}$ $\star$   & $ (\bold{1_{-}},1), (\bold{2_{k}}, 1) $   \\
\hline
 \multirow{2}{*}{$D_{N \in even}$ $\star$}   & $ (\bold{1_{-+}}, 1),  (\bold{1_{--}}, 1), (\bold{2_{k}}, 1) $   \\
   & $ (\bold{1_{+-}}, 1),  (\bold{1_{--}}, 1), (\bold{2_{k}},1) $\\
\hline
$S_{3}$ $\star$  & $ (\bold{1^{\prime}}, 1), (\bold{2},1) $ \\
\hline
$S_{4}$  $\star$   & $ (\bold{1^{\prime}}, 1), (\bold{2}, 1), (\bold{3}, 1) $ \\
\hline
$A_{4}$ $\star$$\star$ & $ (\bold{1^{\prime}}, 2/3), (\bold{1^{\prime\prime}}, -2/3) $   \\
\hline
 $A_{N \ge 5}$   & $ (\text{null}, 0) $ \\
\hline
\multirow{2}{*}{$Q_{N, N/2 \in even}$ $\star$} & $ (\bold{1_{-+}}, 1), (\bold{1_{--}}, 1), (\bold{2_{k_{e}}}, 1) $\\
 & $ (\bold{1_{+-}}, 1), (\bold{1_{--}}, 1), (\bold{2_{k_{e}}}, 1) $ \\
\hline
\multirow{2}{*}{$Q_{N, N/2 \in odd}$ $\star$}  & $  (\bold{1_{+-}}, 1/2), (\bold{1_{-+}}, -1/2), (\bold{1_{--}}, 1), (\bold{2_{k_{e}}}, 1) $   \\
& $\circ$\,\, $(\bold{1_{+-}}, 1), (\bold{1_{-+}}, 1) $\\
\hline
\multirow{2}{*}{$QD_{2N}$ $\star$}  & $(\bold{1_{+-}}, 1), (\bold{1_{--}}, 1), (\bold{2_{k_{o}}}, 1), (\bold{2_{k_{e}}}, 1)$\\
& $ (\bold{1_{-+}}, 1), (\bold{1_{--}}, 1), (\bold{2_{k_{o}}}, 1) $ \\
\hline
 $T_{N}$ $\star$$\star$ & $ (\bold{1_{1}}, 2/3), (\bold{1_{2}}, -2/3) $ \\
\hline
$T^{\prime}$ $\star$$\star$ & $ (\bold{1^{\prime}}, 2/3), (\bold{1^{\prime\prime}}, -2/3), (\bold{2^{\prime}}, -2/3), (\bold{2^{\prime\prime}}, 2/3)$  \\
\hline
\multirow{2}{*}{$\Delta(3N^{2})_{ N/3 \in \mathbb{Z}}$$\star$$\star$} & $ (\bold{1_{k,l}}, 2k/3) $\\
& $ (\bold{1_{k,l}}, 2l/3) $  \\
\hline
$\Delta(3N^{2})_{N/3 \notin \mathbb{Z}}$ $\star$$\star$ & $ (\bold{1_{k}}, 2k/3) $ \\
\hline
$\Delta(6N^{2})_{3N \in \mathbb{Z}}$  $\star$ & $ (\bold{1_{1}}, 1),  (\bold{2_{n}}, 1),  (\bold{3_{1k}}, 1),  (\bold{6_{[k][l]}}, 1) $ \\
\hline
$\Delta(6N^{2})_{3N \notin \mathbb{Z}}$  $\star$ & $ (\bold{1_{1}}, 1),  (\bold{2}, 1),  (\bold{3_{1k}}, 1),  (\bold{6_{[k][l]}}, 1) $  \\
\hline
\multirow{2}{*}{$\Sigma(2N^{2})$ $\star$}  & $\circ$\,\,  $(\bold{1_{-n}}, 1), (\bold{2_{p,q}}, 1)$ \\
&  $(\bold{1_{\pm n}}, 2n/N), (\bold{2_{p,q}}, 2(p+q)/N)$    \\
\hline
\multirow{2}{*}{$\Sigma(3N^{3})$ $\star$$\star$} &    $\circ$\,\, $(\bold{1_{k,l}}, 2k/3)$ \\
&  $(\bold{1_{k,l}}, 2l/N), (\bold{3_{[l][m][n]}},2(l+m+n)/N)$ \\
\hline
\end{tabular}
\caption{Logarithms of basis elements to be employed in \eqref{eq:basisthree} for various finite groups.  For groups occupying two rows, both anomaly constraints implied by the parameter sets must be met simultaneously.  One/two star(s) ($\star$) indicates that the group is subject to $D_{(1)/(2)}$ in \eqref{eq:genfield}-\eqref{eq:genfield2}. NOTES: $\bold{a)}$ The subscript `e/o' indicate `even/odd' $\bold{b)}$ For $\Delta(3N^{2})$, $(k,l) \in \lbrace 0,1,2 \rbrace$ $\bold{c)}$ For $D_{N \in (e,o)}$, $k \in \lbrace 1 ... (N/2 -1,(N-1)/2) \rbrace$ $\bold{d)}$ For $\Delta(6N^{2})$, $n \in \lbrace 1 ... 4 \rbrace$ and $k \in \lbrace 1...N-1 \rbrace$.  For $3N \,(\in, \notin)\, \mathbb{Z}$, there are $(N(N-3)/6,(N^{2}-3N+2)/6)$ sextets. $\bold{e)}$ For $\Sigma(2N^{2})$, $n \in \lbrace 0...N-1 \rbrace$ and there are $N(N-1)/2$ doublets $\bold{f)}$ For $\Sigma(3N^{3})$ there are $3N$ singlets and $N(N^{2}-1)/3$ triplets.}
\label{tab:two}
\end{table}

Consider the case where a non-Abelian discrete symmetry is appended to a single non-Abelian gauge group, as occurs in many $SU(5)$ and $SO(10)$ Grand Unified models.  Respectively denoting the number of fields simultaneously in the $\bold{r}$ and $\bold{d}$ irreps as $\phi_{(\bold{d},\bold{r})}$, the  $D-G-G$ anomaly constraint from \eqref{eq:basisthree} then becomes:
\begin{align}
\label{eq:anommasterGUT}
B^{G}_{(\bold{d},\bold{r})} &= \sum_{\bold{d}} \sum_{\bold{r}} 2\,l(\bold{r}) \cdot \left[ \phi_{(\bold{d},\bold{r})}\right] \cdot \ln x_{\bold{d}} \\
&\equiv \sum_{\bold{d}} K^{G}_{1}(\phi_{\bold{d}}) \cdot \ln x_{\bold{d}} 
\end{align}
where we have defined the field kernel $K^{G}_{1}(\phi_{\bold{d}})$ for a single gauge factor and left its dependence on $\bold{r}$ implicit.\footnote{Note that sums over $f$ are now gone, as implied by the introduction of the parameters $\phi_{(\bold{d},\bold{r})}$ which are by definition $\in \mathbb{Z}^{+}$ (the positive integers including zero).}  The constraint becomes more complex when two gauge symmetries are considered.  Explicitly writing the gauge symmetry factor, $\hat{\bold{r}} \equiv \text{dim}(\bold{r})$, the field kernel becomes
\begin{align}
\nonumber
K^{G}_{2}(\phi_{\bold{d}}) = \sum_{i} 2\, l(\bold{r}_{i}) \cdot \left[ \sum_{j}\hat{\bold{r}}_{j} \cdot \phi_{(\bold{d},\bold{r}_{i},\bold{r}_{j})} \right] 
\end{align}
with the subscripts on the parameters $\phi$ denoting the relevant representations under all three symmetries.  It is understood that the anomaly constraint from $i \leftrightarrow j$ must be satisfied simultaneously and that $\bold{r}_{i=1} \neq \bold{r}_{j=1} = \bold{1}$.  Continuing, the number of subscripts on $\phi$, sums and symmetry factors within the square brackets, and independent discrete anomaly constraints will increase by one for each additional gauge symmetry considered.  

On the other hand, the structure of the analogous $D-\mathcal{G}-\mathcal{G}$ field kernel is universal:
\begin{align}
\label{eq:anommasterGRAV}
K^{\mathcal{G}}(\phi_{\bold{d}}) = \sum_{\lbrace \bold{r} \rbrace}\,2\,\left[\prod_{i=1}^{m}\, \hat{\bold{r}}_{i}\,\right] \phi_{(\bold{d},\lbrace \bold{r} \rbrace)}  
\end{align}
where $m$ denotes the number of symmetries $G_{i}$ in the theory and the sum is over the set of unique gauge symmetry assignments $\lbrace \bold{r} \rbrace \sim \left( \bold{r}_{1},\bold{r}_{2}, ... \bold{r}_{m} \right)$, where (e.g.) $\left( \bold{1}, \bold{2}, \bold{3}, ...  \right) \neq \left( \bold{2}, \bold{1}, \bold{3}, ...  \right)$ and so on.

We now make the observation that the non-Abelian groups catalogued in Table \ref{tab:two} are generically subject to one of two classes of constraints distinguished by the following `field equations':
\begin{align}
\label{eq:genfield}
D_{(1)}:& \,\,\,\,\, \sum_{\bold{d}} K^{(G,\mathcal{G})}(\phi_{\bold{d}}) \overset{!}{=} 2n, \,\,\, n \in \mathbb{Z}^{\pm}\\
\label{eq:genfield2}
D_{(2)}:& \,\,\,\,\, 
\sum_{\bold{d}_{+}} K^{(G,\mathcal{G})}(\phi_{\bold{d}_{+}}) \overset{!}{=} 3n + \sum_{\bold{d}_{-}} K^{(G,\mathcal{G})}(\phi_{\bold{d}_{-}}) 
\end{align}
where the $\bold{d}_{\pm}$  notation indicates irreps with positive or negative $\ln x_{\bold{d}}$, and where free parameters in these basis logarithms  (like $k$ or $l$ for $\Delta(3N^{2})$) are also implied in the field kernels. In Table \ref{tab:two} we indicate symmetries governed by $D_{(1)}$ and $D_{(2)}$ with one or two stars ($\star$), respectively.  Note that satisfying \eqref{eq:genfield}-\eqref{eq:genfield2} is necessary but not sufficient to determine complete anomaly freedom for some groups, as they only have one independent discrete transformation (labeled by a ($\circ$) in Table \ref{tab:two}) subject to \eqref{eq:genfield} or \eqref{eq:genfield2}.

\subsection{Pocket Formulae and Results}
\label{sec:derivatives}
We can now derive a handful of powerful consequences from \eqref{eq:genfield}-\eqref{eq:genfield2} relevant to realistic BSM scenarios:
\begin{enumerate}

\item  \label{itm:grav1D1} Any model subject only to $D_{(1)}$ is free of gravitational anomalies.\footnote{This conclusion is consistent with the observation in \cite{Ishimori:2010au} that $\left[ \mathcal{G} \right]^{2}D$ anomalies are trivially satisfied by $\mathcal{O}(2)$ discrete transformations.}  

\item  \label{itm:1G1D1} Any model subject only to $D_{(1)}$ is free of gauge anomalies if 
\begin{equation}
\label{eq:integer}
l(\bold{r}) \in \mathbb{Z}^{+}\, \forall\, f
\end{equation}
This is the case for (e.g.) $SO(10)$, $E_{6}$, $E_{7}$, $E_{8}$, $F_{4}$, and $G_{2}$ theories, and therefore anomaly cancellation for such theories proceeds automatically, as it does when considering models based on these continuous groups alone (see e.g. \cite{Slansky:1981yr}).

\item Points \ref{itm:grav1D1} and  \ref{itm:1G1D1} are consistent with, and provide concrete examples of, those drawn in \cite{Chen:2015aba} regarding anomaly freedom for non-perfect finite groups, and \eqref{eq:GENconstraint2}-\eqref{eq:CJconstraint2} further imply that a condition for such groups to be generically anomaly free is, in addition to \eqref{eq:integer} for gauge anomalies, given by
\begin{equation}
 \left(\det\left[h_{i}(\bold{d}_{j})\right]\right)^{2} = 1 \,\forall \, \lbrace i, j \rbrace
 \end{equation}
or equivalently for all $C_{i}(\bold{d}_{j})$.  

\item For the special case of $SU(5)$ Grand Unified constructions subject to $D_{(1)}$, fermions in the $\bold{5}$ and $\bold{10}$ (and conjugates) are often the only exceptions to \eqref{eq:integer}. Then the sum of all such fermions in irreps $\bold{d}$ must itself be even,\footnote{Our convention is such that $l(\bold{F})$ is $\frac{1}{2}$ and $1$ for fundamentals $\bold{F}$ in $SU(N)$ and $SO(N)$ groups respectively \cite{Bernard:1977nr,Yamatsu:2015npn}.} \footnote{Fields in conjugate gauge irreps $\bar{\bold{r}}$ are counted within $\phi_{(\bold{d}, \bold{r}_{1},...)}$ such that, e.g., $\phi_{(\bold{d},\bold{5})} \equiv \phi_{(\bold{d},\bold{5})} + \phi_{(\bold{d},\bar{\bold{5}})}$.}
\begin{equation}
\label{eq:SU5simple}
\sum_{\bold{d}} \phi_{(\bold{d},\bold{5})} + 3\,\phi_{(\bold{d},\bold{10})} \overset{!}{=} 2n
\end{equation}
It is easy to extend this to include additional gauge irreps.

\item  \label{itm:2G1D1} For models subject to $D_{(1)}$ and employing multiple gauge symmetries $G$, but possibly non-integer $l(\bold{r})$, anomaly freedom is only determined once the representations $\bold{r}_{i}$ under $G_{i}$ are specified.  As a special but important case, consider an extension to the SM with all chiral fields transforming under the trivial or (anti-)fundamental irreps of the SM gauge groups, as is often the case in BSM flavour and dark matter models.  Anomaly freedom then requires
\begin{align}
\label{eq:SMD1}
\sum_{\bold{d}} \left[ \phi_{(\bold{d},\bold{3},\bold{1})} + \bold{2} \cdot \phi_{(\bold{d},\bold{3},\bold{2})} \right] &\overset{!}{=} 2n\\
\label{eq:SMD1b}
\sum_{\bold{d}} \left[ \phi_{(\bold{d},\bold{2},\bold{1})} + \bold{3} \cdot \phi_{(\bold{d},\bold{2},\bold{3})} \right] &\overset{!}{=} 2n
\end{align}
sourced from $\left[SU(3) \right]^{2} D_{(1)}$ and $\left[SU(2) \right]^{2} D_{(1)}$ triangles, respectively.  Note that in the former case the restriction actually reduces to
\begin{equation}
\label{eq:RHquark}
\sum_{\bold{d}} \phi_{(\bold{d},\bold{3},\bold{1})} \overset{!}{=} 2n
\end{equation}
which normally amounts to counting the number of singlet quarks non-trivially charged under $D$.  Similar considerations can be made for Pati-Salam constructions.  

\item   \label{itm:grav1D2} Any model subject to $D_{(2)}$ and employing one $G$ suffers from gravitational anomalies if \footnote{Not including free parameters like $k$ or $l$ for $\Delta(3N^{2})$...}
\begin{equation}
\label{eq:grav1D2}
\,\,\,\,\,\,\,\,\sum_{\bold{d}_{+}} \sum_{\bold{r}} \hat{\bold{r}} \cdot \phi_{(\bold{d}_{+},\bold{r})} \overset{!}{=} \frac{3}{2} n + \sum_{\bold{d}_{-}}  \sum_{\bold{r}} \hat{\bold{r}} \cdot\phi_{(\bold{d}_{-},\bold{r})}
\end{equation}
is not satisfied, and from gauge anomalies if 
\begin{equation}
\,\,\,\,\,\,\,\,\sum_{\bold{d}_{+}}\sum_{\bold{r}} l(\bold{r}) \cdot \phi_{(\bold{d}_{+},\bold{r})} \overset{!}{=} \frac{3}{2} n + \sum_{\bold{d}_{-}}\sum_{\bold{r}} l(\bold{r}) \cdot\phi_{(\bold{d}_{-},\bold{r})}
\end{equation}
is not satisfied.  For the $T_{N}$ and $A_{4}$ groups there is only one $\bold{d}_{\pm}$ irrep each. 

\item \label{itm:1G1D2} Any model subject to $D_{(2)}$ and employing multiple symmetries $G$ suffers from gravitational anomalies if 
\begin{align}
\nonumber
&\sum_{\bold{d}_{+}}\sum_{\lbrace \bold{r} \rbrace} \left[ \prod _{i=1}^{m} \hat{\bold{r}}_{i} \right] \cdot \phi_{(\bold{d}_{+},\lbrace \bold{r} \rbrace)} \overset{!}{=} \frac{3}{2} n \\
\label{eq:grav2D2}
+ &\sum_{\bold{d}_{-}}\sum_{\lbrace \bold{r} \rbrace}  \left[ \prod _{i=1}^{m} \hat{\bold{r}}_{i} \right] \cdot\phi_{(\bold{d}_{-},\lbrace \bold{r} \rbrace)}
\end{align}
As an obvious point, it should be clear that the sums of \emph{neither} or \emph{both} $\bold{d}_{\pm}$ field kernels must be a multiple of three, a fact that in some instances may be easier to exploit.

\item  \label{itm:2G1D2}  For models employing multiple $G$ and subject to $D_{(2)}$ it is generally easier to determine gauge anomaly freedom by expanding \eqref{eq:genfield2} for the particular discrete symmetry at hand, which normally has (at most) a handful of relevant $\bold{d}$.   The structure of the field kernels will mimic the LHS of \eqref{eq:SMD1}-\eqref{eq:SMD1b}, with $\bold{d} \rightarrow \bold{d}_{\pm}$, for the special case with SM gauge structure and fundamental or trivial irreps.

\end{enumerate}

Of course, if anomalies are encountered, it may still be possible to cancel them with the discrete version of the Green-Schwarz Mechanism \cite{Green:1984sg,Lee:2011dya,Chen:2013dpa}, although the phenomenology of the model may also be altered \cite{Chen:2015aba}.  In the event one also wishes to preserve MSSM type gauge coupling unification, there is the further requirement of `anomaly universality' \cite{Chen:2013dpa,Chen:2012jg,Ibanez:1994ig} which forces \eqref{eq:basisthree} to be equal for all gauge groups $G_{i}$ in the theory, e.g.
\begin{equation}
\label{eq:universal}
B^{(Z,SU(3))}_{(\bold{d},\bold{r})} = B^{(Z,SU(2))}_{(\bold{d},\bold{r})} = \rho \,\, \text{mod} \,\,2 \pi i 
\end{equation}   
for SM constructions (with $Z$ representing an independent Abelian transform of the larger non-Abelian group $D$ and $\rho$ possibly non-zero).  Hence anomaly universality forces the LHS of, e.g.,  \eqref{eq:SMD1}-\eqref{eq:SMD1b} to both be equal modulo two. Similar constraints for other simplified formulae also hold.
\section{Applications}
\label{sec:APPS}
\begin{table}[t]
\renewcommand{\arraystretch}{1.5}
\noindent
\centering
\begin{tabular}{|c|c|c|c|c|}
\hline
$\bold{\text{(1)}}$  & $L_{\mu}$ & $L_{\tau}$ & $l^{c}_{\mu}$ & $l^{c}_{\tau}$   \\
\hline 
$SU(2)$ & $\bold{2}$ & $\bold{2}$ & $\bold{1}$ & $\bold{1}$ \\
\hline
$A_{4}$ & $\bold{1^{\prime}}$ & $\bold{1^{\prime \prime}}$ & $\bold{1^{\prime \prime}}$ & $\bold{1^{ \prime}}$ \\
\hline
\end{tabular}\, \\
\,\,\,
\begin{tabular}{|c|c|c|}
\hline
$\bold{\text{(2)}}$  & $(D_{\mu}, D_{\tau})$ & $(D_{Q_{1}}, D_{Q_{2}})$ \\
 \hline 
 $(SU(2),SU(3))$  & $(\bold{2},\bold{1})$ & $(\bold{2},\bold{3})$   \\
\hline 
$S_{3}$  & $\bold{2}$ & $\bold{2}$  \\
\hline
\end{tabular}\, \\
\,\,\,
\begin{tabular}{|c|c|}
\hline
$\bold{\text{(3)}}$  & $(D_{\mu}, D_{\tau})$   \\
 \hline 
 $SU(2)$  & $\bold{2}$    \\
\hline 
$D_{4}$  & $\bold{2}$   \\
\hline
\end{tabular}\,
\begin{tabular}{|c|c|}
\hline
$\bold{\text{(4)}}$  & $t^{c}$  \\
\hline
$SU(3)$  & $\bold{\bar{3}}$ \\
\hline
$\Delta(6N^{2})$ & $\bold{1_{1}}$\\
\hline
\end{tabular}\,
\begin{tabular}{|c|c|}
\hline
$\bold{\text{(5)}}$  & $F$  \\
\hline
$SU(5)$ & $\bold{\bar{5}}$\\
\hline
$S_{4}$ & $\bold{3}$  \\
\hline
\end{tabular}
\\
\,\,\,
\begin{tabular}{|c|c|c|c|c|c|c|c|c|c|c|c|}
\hline
$\bold{\text{(6)}}$  & $H_{24}$ & $\Lambda_{24}$ & $X_{5}$ & $X_{6}$ & $X_{8}$ & $X_{9}$ & $X_{10}$ & $Z_{1,2}$ & $Z_{3}$ & $\Upsilon_{i}$ & $\Upsilon_{j}$ \\
\hline
$SU(5)$ & $\bold{24}$ & $\bold{24}$ & $\bold{\bar{5}}$ & $\bold{5}$ & $\bold{5}$ & $\bold{\bar{5}}$ & $\bold{5}$ & $\bold{24}$ & $\bold{24}$ & $\bold{24}$ & $\bold{24}$\\
\hline
$A_{4}$ & $\bold{1^{\prime}}$ & $\bold{1^{\prime}}$ & $\bold{1^{\prime \prime}}$ & $\bold{1^{\prime}}$  & $\bold{1^{\prime \prime}} $ &  $\bold{1^{\prime}}$ &  $\bold{1^{\prime}}$ & $\bold{1}^{\prime \prime}$ & $\bold{1}^{\prime}$ & $\bold{1}^{\prime}$ & $\bold{1}^{\prime \prime}$\\
\hline
\end{tabular}
\caption{$\bold{\text{(1)}}$:  The relevant field and symmetry content from the $A_{4}$ dark matter model of \cite{Hirsch:2010ru}. $\bold{\text{(2)}}$:  The same for the $S_{3}$ flavour model of \cite{Feruglio:2007hi}.  $\bold{\text{(3)}}$:  The same for the $D_{4}$ leptonic flavour model of \cite{Grimus:2003kq}. $\bold{\text{(4)}}$:   The same for the $\Delta(6N^{2})$ quark flavour model of \cite{Ishimori:2014jwa}. $\bold{\text{(5)}}$:  The same for the $S_{4}$ GUT model of \cite{Meloni:2011fx}. $\bold{\text{(6)}}$: The same for the $A_{4}$ GUT model of \cite{Bjorkeroth:2015ora}.  Here $i = \lbrace 1,3, 5, 7 \rbrace$ and $j = \lbrace 2, 4, 6, 8 \rbrace$.}
\label{tab:models}
\end{table}

We apply the constraints found in Section \ref{sec:derivatives} to a host of models representative of common BSM symmetry environments.  We  present only the field and symmetry content of the models required to calculate the non-Abelian discrete anomalies, and in each case we only probe the  `easiest' simplified constraints from Section \ref{sec:derivatives} until we determine (if) the model is anomalous.  We do not address the possibility that effective theories may receive anomaly contributions from additional light states in the UV, thereby changing the low-energy conclusions.

\subsection{$A_{4}$ Dark Matter Model of \cite{Hirsch:2010ru}}
\label{sec:A4dark}
All chiral fermions in this model are in the trivial or fundamental irreps of (at least one of) the SM gauge groups.
We observe from Table \ref{tab:models} that $\phi_{(\bold{1^{\prime}},\bold{2},\bold{1})} = \phi_{(\bold{1^{\prime \prime}},\bold{2},\bold{1})} = \phi_{(\bold{1^{\prime \prime}},\bold{1},\bold{1})} =\phi_{(\bold{1^{\prime}},\bold{1},\bold{1})} = 1$ and zero for all other entries. This gives field kernels of $K^{(G,\mathcal{G})}(\phi_{\bold{d}_{+}}) = K^{(G,\mathcal{G})}(\phi_{\bold{d}_{-}})$ trivially satisfying \eqref{eq:genfield2} for both gauge and gravitational constraints.  The model is therefore anomaly free.

\subsection{$S_{3}$ Flavour Model of \cite{Feruglio:2007hi}}
\label{sec:flavS3}
This model is subject to \eqref{eq:SMD1}-\eqref{eq:SMD1b}.  Considering $\left[SU(2) \right]^{2} D_{(1)}$ anomalies, from Table \ref{tab:models} we count $\phi_{(\bold{2},\bold{2},\bold{1})} = \phi_{(\bold{2},\bold{2},\bold{3})} = 1$ and zero for all other contributions, satisfying \eqref{eq:SMD1b}.  For  $\left[SU(3) \right]^{2} D_{(1)}$, we find a lone contribution from $\phi_{(\bold{2},\bold{3},\bold{2})} = 1$, which also satisfies \eqref{eq:SMD1}.  As $D_{(1)}$ symmetries do not suffer from gravitational anomalies, we conclude that this model is anomaly free.

\subsection{$D_{4}$ Leptonic Flavour Model of \cite{Grimus:2003kq}}
\label{sec:A4dark}
This $D_{4}$ flavour model provides a final example of the power of  \eqref{eq:SMD1}-\eqref{eq:SMD1b}. From Table \ref{tab:models} we find that the $SU(2)_{L}$ doublets provide the only anomalous contributions, giving $\phi_{(\bold{2},\bold{2},\bold{1})} = 1$.  It is clear that \eqref{eq:SMD1b} can never be realized and thus the model suffers from $\left[SU(2) \right]^{2} D_{(1)}$ anomalies, a conclusion consistent with \cite{Araki:2008ek}.

\subsection{$\Delta(6N^{2})$ Quark Flavour Model of \cite{Ishimori:2014jwa}}
\label{sec:6N2flav}
We take the symmetry assignments of additional flavons and driving superfields to be trivial under the SM gauge group, as is standard.  However, from Table \ref{tab:models} we immediately see that $\phi_{(\bold{1}_{1},\bold{3},\bold{1})} = 1$ and that by virtue of \eqref{eq:RHquark} the model suffers from $\left[SU(3) \right]^{2} D_{(1)}$ anomalies.  

\subsection{$SU(5) \times S_{4}$ Grand Unified Model of \cite{Meloni:2011fx}}
\label{sec:grandS4}
We count that $\phi_{(\bold{3}, \bold{5})} = 1$ and zero for all other parameters relevant to gauge constraints, so \eqref{eq:SU5simple} can never be realized and hence the model suffers from $\left[SU(5) \right]^{2}D_{(1)}$ anomalies.  Note that this model is extra-dimensional, with the field $F$ presented in Table \ref{tab:models} living on the brane.

\subsection{$SU(5) \times A_{4}$ Grand Unified Model of \cite{Bjorkeroth:2015ora}}
\label{sec:grandA4}
The relevant field content of \cite{Bjorkeroth:2015ora} sits in both the fundamentals and adjoints of $SU(5)$: $\phi_{(\bold{1^{\prime}},\bold{5})} = 3$, $\phi_{(\bold{1^{\prime \prime}},\bold{5})} = 2$, $\phi_{(\bold{1^{\prime}},\bold{24})} = 7$, and  $\phi_{(\bold{1^{\prime \prime}},\bold{24})} = 6$.  The quickest Type-2 constraint comes from \eqref{eq:grav1D2}, which gives:
\begin{equation}
24\cdot7 + 5\cdot 3 \neq \frac{3}{2} n + 5\cdot 2 + 24 \cdot 6
\end{equation}
implying that the model suffers from gravitational anomalies.

\section{Conclusion}
\label{sec:conclusions}
We have shown that models employing non-Abelian discrete symmetries are typically subject to one of two classes of anomaly constraint, the first restricting the sum of fields charged under $G$ and $D$ to be even, and the second restricting them to be a multiple of three, upon accounting for all relevant gauge symmetry factors.  These simple equations have powerful implications in realistic BSM environments, especially Grand Unified scenarios.  Of course, specificity is always limited by scope, and hence it would be interesting to study the derivatives of our generic formulae when additional theoretical or phenomenological considerations are imposed on the gauge and/or discrete symmetry structure of a theory; it is likely that in specific model building environments (e.g. dark matter) even more powerful constraints on acceptable field contents arise. 

\section{Acknowledgements}
I am grateful to Sven Krippendorf for inspiring conversations at the beginning of this work, to Graham Ross for many important insights, and to both of them for their review of the manuscript.  I acknowledge research and travel support from DESY, and thank the University of Oxford for hospitality during the completion of portions of this project.

\end{document}